\documentclass[a4paper]{jpconf}
\usepackage{graphicx}
\usepackage{iwsmihead}
\usepackage{amssymb}

\newcommand{\bb}{\mbox{\boldmath{$b$}}}

\newcommand{\bx}{\mbox{\boldmath{$x$}}}
\newcommand{\brx}{\mbox{\bf x}}

\newcommand{\bc}{\mbox{\boldmath{$c$}}}
\newcommand{\bzero}{\mbox{\boldmath{$0$}}}
\newcommand{\by}{\mbox{\boldmath{$y$}}}

\newcommand{\mR}{\mathbb{R}}

\newcommand{\paritycheck}{{\left \langle l_1 l_2 \ldots l_k \right \rangle}}

\begin{document}
\title{How could the replica method improve accuracy of 
performance assessment of channel coding?}

\author{Yoshiyuki Kabashima}

\address{Department of Computational Intelligence and 
Systems Science \\
Tokyo Institute of technology, 
Yokohama 226-8502, Japan}

\ead{kaba@dis.titech.ac.jp}

\begin{abstract}
We explore the relation between 
the techniques of statistical mechanics and 
information theory for assessing 
the performance of channel coding.	
We base our study on a framework developed by Gallager 
in {\em IEEE Trans. Inform. Theory} {\bf 11}, 3 (1965), 
where the minimum decoding error probability 
is upper-bounded by an average 
of a generalized Chernoff's bound over 
a code ensemble. 
We show that the resulting bound in the framework
can be directly assessed by the replica method, 
which has been developed in statistical mechanics of disordered systems, 
whereas in Gallager's original methodology further replacement by another bound 
utilizing Jensen's inequality is necessary. 
Our approach associates 
a seemingly {\em ad hoc} restriction with respect to 
an adjustable parameter for optimizing the bound with 
a phase transition between two replica symmetric solutions, 
and can improve the accuracy of performance assessments 
of general code ensembles including low density 
parity check codes, 
although its mathematical justification is still open. 
\end{abstract}

\section{Introduction}
In the last few decades, much attention has been paid 
to the similarities between statistical mechanics and information theory. 
In general, inference or search problems that arise
in research on communication, inference, learning, combinatorics 
and other information theory fields can be treated by regarding the system as a virtual spin system 
subject to disordered 
interactions \cite{Nishimori2001,WatkinRauBiehl1992,Engel2001}. 
In this way, 
problems in information theory have been successfully analyzed
utilizing methods developed in statistical mechanics 
\cite{KabashimaSaad1998,KabashimaSaad1999,MontanariSourlas2001,
Tanaka2002,Kabashima2003a}, 
and vice versa \cite{MezardParisi2002,Kabashima2003b}. 

This research trend has shown that the similarities between the 
two fields are not limited to the structure of 
problems but also apply to analysis techniques. 
However, because 
the development histories of the two frameworks have been relatively independent, there are still barriers 
which may hinder further expansion and deepening of this 
promising interdisciplinary research field. 
In order to overcome possible obstacles, 
it is of great importance to investigate the methodological relations between the two 
fields. This article is written under this motivation. 
More precisely, we explore the similarities
and differences between the techniques of statistical mechanics and 
information theory in analyzing channel coding
(or error correcting codes). 

This article is organized as follows. 
In sections 2 and 3, we briefly 
review a standard framework of (classical) channel coding 
and a conventional methodology for assessing its performance, 
which was developed by Gallager in \cite{Gallager1965}. 
Sections 4 and 5 are the main parts of the current article. 
In section 4, we reconsider the channel coding problem 
by applying the replica method developed in statistical mechanics. 
Using the replica method makes it possible to 
avoid applying Jensen's inequality, which is required 
in the original methodology. This offers a novel 
interpretation of the origin of a superficially 
{\em ad hoc} restriction with respect to an adjustable 
parameter for tightening the upper-bound of 
the minimum decoding error probability that 
appears in the conventional approach. Applying the replica method does not 
change the assessed performance, 
though it can 
improve the accuracy of the performance assessment for general 
code ensembles, including low-density parity-check codes, 
as shown in section 5. 
The final section, section 6, is devoted to a summary and discussion.

\section{Framework of channel coding}
Consider 
a message $m \in \{1,2,\ldots,2^K\}$ 
transmitted to a receiver through a classical noisy channel. 
For this purpose, $m$ is, in general, mapped to a codeword 
of $N$ dimension $\bx_m =(x_{m1},x_{m2},\ldots,x_{mN})
\in \{0,1\}^N$ prior to the transmission. 
The mapping of $m \to \bx_m$
$(m=1,2,\ldots,2^K)$ can equivalently 
be expressed as ${\cal C}=\{\bx_1,\bx_2,\ldots,\bx_{2^K}\}$ and is 
termed a {\em channel coding} or simply a (channel) {\em code}. 

The receiver must infer the original 
message $m$ from the received degraded codeword
of $N$ dimension $\by=(y_1,y_2,\ldots,y_N)$. 
For simplicity, we assume a memoryless channel with the degradation process
modeled by a conditional probability
$P(\by|\bx_m)=\prod_{l=1}^N P(y_l|x_{ml})$. 
We further assume that the message $m$ is encoded by a method of
{\em source coding} such that it is equally generated with a probability 
of $2^{-K}$, which is preferred for enhancing 
communication performance \footnote{In the conventional argument 
of information theory, channel coding is examined 
independently of source coding without assuming a prior distribution 
of messages. However, we here assume that messages are uniformly 
distributed {\em a priori} as a result of source coding 
in order to emphasize the optimality of the maximum-likelihood decoding.}. 
Under these assumptions, Bayesian theory
indicates that for a given code ${\cal C}$, 
the maximum likelihood (ML) decoding 
\begin{eqnarray}
\widehat{m}(\by)=\mathop{\rm argmax}_{s \in \{1,2,\ldots,2^K\}}
\left \{P(\by|\bx_s) \right \}, 
\label{MLdecoding}
\end{eqnarray}
minimizes the probability of decoding error
\begin{eqnarray}
P_{e}({\cal C})=2^{-K}\sum_{m,\by}
P(\by|\bx_m)\Delta_{\rm ML}(m,\by),
\label{errorprob}
\end{eqnarray}
where ${\rm argmax} \{ \cdots \}$ denotes the argument
that maximizes $\cdots$ and $\Delta_{\rm ML}(m,\by)=1$ if the original message
$m$ is not correctly retrieved by equation (\ref{MLdecoding}) 
for a given $\by$ and $\Delta_{\rm ML}(m,\by)=0$ otherwise. 
In the following, we address the problem of assessing how small 
a $P_{e}({\cal C})$ is achievable by selecting 
the optimal ${\cal C}$ among a given code ensemble. 

\section{Conventional scheme for analyzing channel coding}
\subsection{Generalized Chernoff's bound}
As $\Delta_{\rm ML}(m,\by)$
depends on $m$ and $\by$ in a highly nonlinear manner, 
direct evaluation of equation (\ref{errorprob}) is 
difficult. In order to avoid this difficulty, 
several techniques for upper-bounding this function have 
been developed in conventional information theory
\cite{Gallager1965,Gallager1968,ViterbiOmura1972}. 

The inequality
\begin{eqnarray}
\Delta_{\rm ML}(m,\by) \le \left (\sum_{s \ne m} 
\left (\frac{P(\by|\bx_s)}{P(\by|\bx_m)} \right )^\lambda \right )^\rho, 
\label{lambda_rho}
\end{eqnarray}
which holds for $\forall{\lambda} > 0$ and $\forall{\rho} > 0$, is 
key for this purpose. This 
is validated as follows. The right hand side is non-negative and 
therefore satisfies the inequality if $\Delta_{\rm ML}(m,\by)=0$. 
If $\Delta_{\rm ML}(m,\by)=1$, there exists at least one message $s$
for which $P(\by|\bx_s) \ge P(\by|\bx_m)$. This means that
for such a message, $\left (P(\by|\bx_s)/P(\by|\bx_m) \right )^\lambda > 1$
holds in the summation of the right hand side of equation 
(\ref{lambda_rho}) since $\lambda > 0$ ensures the fraction 
is greater than unity and therefore the summation is greater than unity 
as all terms are non-negative.
$\rho > 0$ also ensures the inequality is valid. 

Substituting equation (\ref{lambda_rho}) into 
equation (\ref{errorprob}) yields a generalized Chernoff's bound 
\begin{eqnarray}
P_e({\cal C}) \le 2^{-K}
\sum_{m,\by} P(\by|\bx_m) \left (\sum_{s \ne m} 
\left (\frac{P(\by|\bx_s)}{P(\by|\bx_m)} \right )^\lambda
\right )^\rho, 
\label{orig_bound}
\end{eqnarray}
which holds for $\forall{\lambda}>0$ and $\forall{\rho} > 0$. 
This indicates that the accuracy of the upper-bound can be 
improved by minimizing the right hand side with respect to 
these parameters. 

\subsection{Ensemble average as an upper-bound for the minimum}
Unfortunately, direct minimization of the right hand side
of equation (\ref{orig_bound}) is non-trivial due to 
the complicated dependence on ${\cal C}$. However, the 
expression can still be useful for assessing 
the minimum error probability 
among all possible codes,  
$P_e={\rm min}_{{\cal C} \in \{\mbox{all codes}\}} \{P_e({\cal C})\}$, 
for classical channels. 

For this purpose, we introduce an ensemble of all codes
${\cal Q}({\cal C})=\prod_{s=1}^{2^K}Q(\bx_s)$, 
where $Q(\bx)$ is an identical distribution for generating 
codewords $\bx_1, \bx_2,\ldots,\bx_{2^K}$ independently. 
Averaging equation (\ref{orig_bound}) with respect to ${\cal Q}({\cal C})$
gives an upper-bound of $P_e$ as
\begin{eqnarray}
P_e &\le& \overline{P_e({\cal C})}
\le \sum_{{\cal C}\in \{\mbox{all codes}\} }{\cal Q}({\cal C})
\left (2^{-K}
\sum_{m,\by} P(\by|\bx_m) \left (\sum_{s \ne m} 
\left (\frac{P(\by|\bx_s)}{P(\by|\bx_m)} \right )^\lambda
\right )^\rho \right ) \cr
&=&2^{-K} \sum_{\by} \left (\sum_{m=1}^{2^K} \sum_{\bx_m}Q(\bx_m)
P(\by|\bx_m)^{1-\lambda\rho} \right )
\sum_{{\cal C} \backslash \bx_m}
\prod_{s \ne m}^{2^K}Q(\bx_s) 
\left (\sum_{s \ne m}P(\by|\bx_s)^\lambda \right )^\rho, 
\label{second_bound}
\end{eqnarray}
due to the fact that the minimum value over a given ensemble is 
always smaller than the average over the ensemble. 
Here, $\overline{\cdots}$ represents the average over a code ensemble
${\cal Q}({\cal C})$ and 
${\cal C} \backslash \bx_m$ denotes a subset of ${\cal C}
=\{\bx_1,\bx_2,\ldots,\bx_{2^K}\}$
from which only $\bx_m$ is excluded. 

\subsection{Jensen's inequality and random coding exponent}
Equation (\ref{second_bound}) is still difficult to assess 
for large $K$ because the right hand side involves 
the fractional moment of a sum of exponentially many terms
$\sum_{{\cal C} \backslash \bx_m}
\prod_{s \ne m}^{2^K}Q(\bx_s) 
\left (\sum_{s \ne m}P(\by|\bx_s)^\lambda \right )^\rho$, 
the direct and rigorous evaluation of which 
requires an exponentially large computational cost even while 
the code ensemble is factorizable with respect to codewords. 
Jensen's inequality
\begin{eqnarray}
&& \sum_{{\cal C} \backslash \bx_m}
\prod_{s \ne m}^{2^K}Q(\bx_s) 
\left (\sum_{s \ne m}P(\by|\bx_s)^\lambda \right )^\rho 
\le \left (\sum_{{\cal C} \backslash \bx_m}
\prod_{s \ne m}^{2^K}Q(\bx_s)
\sum_{s \ne m}P(\by|\bx_s)^\lambda \right )^\rho \cr
&&=(2^K-1)^\rho \left (\sum_{\bx}Q(\bx)P(\by|\bx)^{\lambda}  \right)^\rho
\le 2^{\rho K}\left (\sum_{\bx}Q(\bx) P(\by|\bx)^{\lambda}  \right)^\rho, 
\label{Jensen}
\end{eqnarray}
which holds for $0 < \rho \le 1$, is a standard technique of
information theory to overcome this difficulty. 
Plugging this into equation (\ref{second_bound}), 
in conjunction with an additional restriction $\rho \le 1$, 
we obtain the expression 
\begin{eqnarray}
P_e &\le& \overline{P_e({\cal C})} \le 2^{\rho K}
\sum_{\by}
\left(\sum_{\bx}Q(\bx)P(\by|\bx)^{1-\rho \lambda} \right )
\left (\sum_{\bx}Q(\bx)P(\by|\bx)^\lambda \right )^\rho, 
\label{rholambdafinal}
\end{eqnarray}
$(0 \le \rho \le 1)$, where 
$2^{-K} \sum_{m=1}^{2^K} \sum_{\bx_m}Q(\bx_m)
P(\by|\bx_m)^{1-\lambda\rho} 
=\sum_{\bx}Q(\bx)P(\by|\bx)^{1-\lambda\rho} $ is utilized and 
the trivial case $\rho=0$ is included. 

For any given $0\le \rho \le 1$, the upper-bound of 
equation (\ref{rholambdafinal}) is generally minimized by 
$\lambda=1/(1+\rho)$, as assumed in Gallager's paper \cite{Gallager1965}. 
The computational difficulty for assessing
equation (\ref{rholambdafinal}) is resolved
for memoryless channels $P(\by|\bx)=\prod_{l=1}^N P(y_l|x_l)$
by assuming factorizable distributions $Q(\bx)=\prod_{l=1}^NQ(x_l)$. 
This assumption naturally indicates that the upper-bound depends
exponentially on the code length $N$ 
as $P_e \le \exp \left [-N\left (-\rho R +E_0(\rho,Q) \right )
\right ]$, where $R=K/N$ and 
\begin{eqnarray}
E_0(\rho,Q)=-\ln \left [\sum_{y}\left (\sum_{x}
Q(x)P(y|x)^{\frac{1}{1+\rho}}\right )^{1+\rho} \right ], 
\label{Gallgaer_function}
\end{eqnarray}
are often termed the code rate and Gallager function, respectively. 
This means that if $N$ is sufficiently large and 
the random coding exponent
\begin{eqnarray}
E_r(R)=\mathop{\rm max}_{0 \le \rho \le 1, Q}
\left \{-\rho R\ln 2+E_0(\rho,Q) \right \}, 
\label{randomcodingbound}
\end{eqnarray}
is positive for a given $R$, 
there exists a code with a decoding error probability 
smaller than an arbitrary positive number. 
For a fixed $Q(x)$, $E_0(\rho,Q)$ is a convex upward function satisfying 
$E_0(\rho=0,Q)=0$ and 
\begin{eqnarray}
\left . 
\frac{\partial}{\partial \rho} E_0(\rho,Q)
\right |_{\rho=0}=\sum_{y,x}Q(x)P(y|x)\ln \frac{P(y|x)}{\sum_{x}Q(x)P(y|x)}
\equiv \ln 2 \times I(Q), 
\label{mutual}
\end{eqnarray}
where $I(Q)$ represents the mutual information between $x$ and $y$
(in bits). 
This implies that the critical rate $R_c$ below which 
$E_r(R)$ becomes positive is given by $\rho=0$ as 
\begin{eqnarray}
R_c=\mathop{\rm max}_{Q} \{I(Q)\}, 
\label{capacity}
\end{eqnarray}
which agrees with the definition of the channel capacity \cite{Shannon1948}. 

\begin{figure}[t]
\setlength{\unitlength}{1mm}
\begin{center}
\includegraphics[width=150mm]{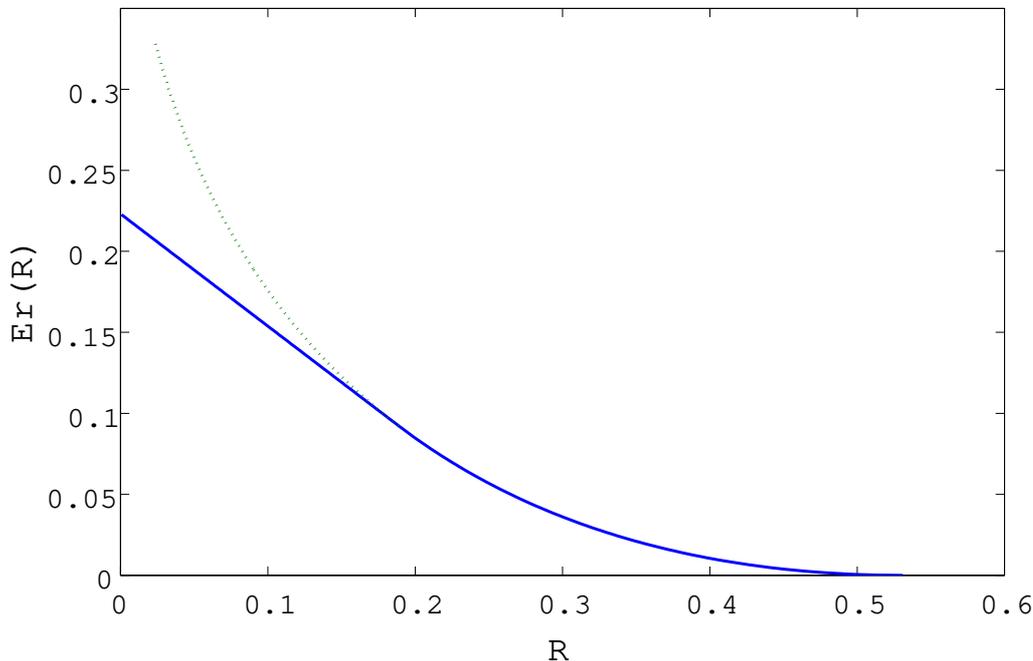}
\end{center}
\caption{Random coding exponent $E_r(R)$ for BSC for a crossover rate 
$p=0.1$. $E_r(R)$ (solid curve) becomes positive 
for $R< R_c(\simeq 0.531)$.
The functional form of $E_r(R)$ for $R< R_b(\simeq 0.189)$
differs from that for $R_b \le R \le R_c$. 
The broken curve represents the value of the upper-bound exponent 
that is maximized without the restriction $\rho \le 1$.  
}
\label{BSCfig}
\end{figure}

As $R$ is reduced from $R_c$, the value of $\rho$ that 
optimizes the right hand side of equation (\ref{randomcodingbound}) 
increases and reaches $\rho=1$ at a certain rate $R_b$. 
Below $R_b$, equation (\ref{randomcodingbound}) is always optimized
at the boundary $\rho=1$. Figure \ref{BSCfig} shows an example of 
$E_r(R)$ for the binary symmetric channel (BSC), which is characterized 
by a crossover rate of $0\le p\le 1$ as $P(1|0)=P(0|1)=p$ and $P(1|1)=P(0|0)=1-p$ 
for binary alphabets $x,y \in \{0,1\}$.

$E_r(R)$ characterizes an upper-bound of a typical decoding error probability 
of randomly constructed codes. However, surprisingly enough, 
it is known that for certain classes of channels, 
$E_r(R)$ represents the performance of the best codes
at the level of exponent for a relatively high code rate region 
$R\ge R_a$, which contains $R=R_b$, 
since $E_r(R)$ agrees with the exponent of a lower bound 
of the best possible code  \cite{Burnashev2005}. 
This is far from trivial because the restriction $\rho \le 1$, 
which governs $E_r(R)$ of $R\le R_b$, is introduced 
in an {\em ad hoc} manner 
when employing Jensen's inequality in the above methodology. 

\section{Performance assessment by the replica method}
\subsection{Expanding the upper-bound for $\rho=1,2,\ldots$}
In order to clarify the origin of the superficially artificial 
restriction $\rho \le 1$, we evaluate the exponent without 
using Jensen's inequality. For this purpose, we
assess the right hand side of equation (\ref{second_bound})
analytically, continuing the expressions obtained for $\rho=1,2,\ldots$
to $\rho \in \mR$. This is often termed the {\em replica method}
\cite{MezardParisiVirasolo1987,Dotzenko2001}. 

For the current problem, the first step of the replica method is 
to evaluate the expression
\begin{eqnarray}
\sum_{{\cal C} \backslash \bx_m}
\!\!\prod_{s \ne m}^{2^K} \! Q(\bx_s)\!  
\left ( \! \sum_{s \ne m}P(\by|\bx_s)^\lambda \! \right )^\rho
\!\! &=&\!\! \sum_{\{s^a\}_{a=1}^\rho}\! \prod_{\tau \ne m}^{2^K}\!
\left (\!\sum_{\bx_\tau}Q(\bx_\tau)P(\by|\bx_\tau)^{\lambda \sum_{a=1}^\rho 
\delta(s^a,\tau)} \!\right ) \cr
&=&\sum_{(i_1,i_2,\ldots,i_\rho)}
\!\!\!
{\cal W}(i_1,i_2,\ldots,i_\rho)
\prod_{t=1}^\rho
\! \left ( \! \sum_{\bx}Q(\bx)P(\by|\bx)^{\lambda t} \! \right )^{i_t}
\!\! , 
\label{replica_expansion}
\end{eqnarray}
analytically for $\rho=1,2,\ldots$, 
where $\delta(x,y)=1$ for $x=y$ and vanishes otherwise,
and ${\cal W}(i_1,i_2,\ldots,i_\rho)$ is the number of 
ways of partitioning $\rho$ replica messages $s^1,s^2,\ldots,s^\rho$
to $i_1$ states (out of $\tau=1,2,\ldots,2^K$ except for $\tau=m$) by one, 
to $i_2$ states by two, $\ldots$ and to $i_\rho$ states by $\rho$. 
Obviously, ${\cal W}(i_1,i_2,\ldots,i_\rho)=0$ unless
$\sum_{t=1}^\rho i_t t=\rho$. 
It is worth noting that the expression 
of the right hand side is valid only for $\rho=1,2,\ldots$. 

\subsection{Saddle point assessment under the replica symmetric ansatz}
Exactly evaluating equation (\ref{replica_expansion})
is difficult for large $K=NR$. 
However, in many systems, quantities of this kind scale 
exponentially with respect to $N$, which implies that 
the exponent characterizing the exponential dependence
can be accurately evaluated by the ``saddle point method''
with respect to the partition of $\rho$, $(i_1,i_2,\ldots,i_\rho)$, 
under an appropriate assumption of the symmetry underlying 
the objective system in the limit of $N \to \infty$. 
The replica symmetry, for which 
equation (\ref{replica_expansion}) 
is invariant under any permutation of the replica indices $a=1,2,\ldots,\rho$, 
is critical for the current evaluation. This implies that 
it is natural to assume that, for large $N$, the final expression of 
equation (\ref{replica_expansion}) is dominated by a {\em single }
term possessing the same symmetry, which yields 
the following two types of replica symmetric (RS) solutions:
\begin{itemize}
\item {\bf RS1}: Dominated by 
$(i_1,i_2,\ldots,i_\rho)=(\rho,0,\ldots,0)$, giving 
\begin{eqnarray}
\sum_{{\cal C} \backslash \bx_m}
\prod_{s \ne m}^{2^K}Q(\bx_s) 
\left (\sum_{s \ne m}P(\by|\bx_s)^\lambda \right )^\rho
&\simeq& {\cal W}(\rho,0,\ldots,0)
\left (\sum_{\bx}Q(\bx)P(\by|\bx)^\lambda \right )^\rho \cr
&\simeq& 2^{N \rho R} \left (\sum_{\bx}Q(\bx)P(\by|\bx)^\lambda \right )^\rho. 
\label{RS1}
\end{eqnarray}
\item {\bf RS2}: Dominated by 
$(i_1,i_2,\ldots,i_\rho)=(0,0,\ldots,1)$, giving
\begin{eqnarray}
\sum_{{\cal C} \backslash \bx_m}
\prod_{s \ne m}^{2^K}Q(\bx_s) 
\left (\sum_{s \ne m}P(\by|\bx_s)^\lambda \right )^\rho
&\simeq &
{\cal W}(0,0,\ldots,1) 
\left (\sum_{\bx}Q(\bx)P(\by|\bx)^{\lambda\rho }\right )^1 \cr
&\simeq & 2^{NR} \sum_{\bx}Q(\bx)P(\by|\bx)^{\lambda\rho }. 
\label{RS2}
\end{eqnarray}
\end{itemize}
Plugging these into the final expression of equation (\ref{second_bound}), 
in conjunction with $P(\by|\bx)=\prod_{l=1}^N P(y_l|x_l)$ and 
$Q(\bx)=\prod_{l=1}^N Q(x_l)$, gives the exponents 
\begin{eqnarray}
E_{\rm RS1}(\rho,\lambda,Q,R)=-\rho R \ln 2 - \ln \left [ \sum_{y}
\left (\sum_{x}Q(x)P(y|x)^{1-\lambda\rho} \right)
\left (\sum_{x}Q(x)P(y|x)^{\lambda} \right)^\rho \right ], 
\label{RS1exponent}
\end{eqnarray}
and 
\begin{eqnarray}
E_{\rm RS2}(\rho,\lambda,Q,R)=-R \ln 2 - \ln \left [ \sum_{y}
\left (\sum_{x}Q(x)P(y|x)^{1-\lambda\rho} \right)
\left (\sum_{x}Q(x)P(y|x)^{\lambda \rho} \right)\right ], 
\label{RS2exponent}
\end{eqnarray}
where the suffixes RS1 and RS2 correspond to equations (\ref{RS1}) and 
(\ref{RS2}), respectively, 
as two candidates of the exponent
$E(\rho,\lambda,Q,R)$ for upper-bounding the minimum decoding 
error probability as $P_e \le \exp \left [-N E(\rho,\lambda,Q,R) \right ]$.

\subsection{Phase transition between RS solutions: 
origin of the restriction $\rho \le 1$}
Although we have so far assumed that $\rho$ is a natural number, 
both the functional forms of the 
saddle point solutions, (\ref{RS1exponent}) and (\ref{RS2exponent}), 
can be defined over $\rho \in \mR$. 
Therefore, we analytically continue these expressions 
from $\rho=1,2,\ldots$ to $\rho \in \mR$, and select the relevant 
solution for each set of $(\rho,\lambda,Q,R)$ 
in order to obtain the correct upper-bound 
exponent $E(\rho,\lambda,Q,R)$.  
This is the second step of the replica method. 

For $\rho=1,2,\ldots$ and sufficiently large $N$, this can be carried out 
by selecting the solution of the lesser exponent value. 
Unfortunately, as yet a mathematically 
justified general guideline for selection of the relevant 
solution for $\rho \le 1$ has not been determined. Such a guideline is 
necessary for determining the channel capacity by assessment at $\rho=0$. 
However, there is an empirical criterion for this purpose, 
which is indicated by the analysis of exactly solvable 
models \cite{OguraKabashima2005}. 
In the current case, this means that for fixed $\lambda,Q$ and $R$ 
we should choose the solution for which the 
partial derivative with respect to $\rho$ at $\rho=1$, 
$\left . (\partial /\partial \rho )
E_{\rm RS1}(\rho,\lambda,Q,R) \right |_{\rho=1}$
or $\left . (\partial /\partial \rho )
E_{\rm RS2}(\rho,\lambda,Q,R) \right |_{\rho=1}$, is lesser, 
as the relevant solution for $\rho \le 1$. 
This criterion implies that $E_{\rm RS1}(\rho,\lambda,Q,R)$ 
should be chosen to
provide the tightest bound $E_{\rm replica}(R)=
\mathop{\rm max}_{0 \le \rho,0 \le \lambda,Q}
\left \{E(\rho,\lambda,Q,R) \right \}$ 
for relatively large $R$, which yields the expression 
\begin{eqnarray}
E_{\rm replica}(R)&=&\mathop{\rm max}_{0 \le \rho,0 \le \lambda,Q}
\left \{E_{\rm RS1}(\rho,\lambda,Q,R) \right \} \cr
&=& \mathop{\rm max}_{0 \le \rho,Q}\left \{
-\rho R \ln 2 -
\ln \left [ \sum_{y}
\left (\sum_{x}Q(x)P(y|x)^{\frac{1}{1+\rho}} \right)^{1+\rho}
\right ] \right \}. 
\label{Ereplica1}
\end{eqnarray}
As $R$ is reduced from $R=R_c$, below which equation (\ref{Ereplica1}) 
becomes positive, the value of $\rho$ that maximizes the right hand 
side of equation (\ref{Ereplica1}) increases from $\rho=0$, keeping the
relation $\lambda=1/(1+\rho)$ at the maximum point. 
When $R$ reaches $R_b$, the optimal value of $\rho$ becomes unity
and $\lambda=1/2$, for which 
\begin{eqnarray}
&&\left . \frac{\partial }{\partial \rho }
E_{\rm RS1}(\rho,\lambda,Q,R) \right |_{(\rho,\lambda,R)=(1,1/2,R_b)}
-\left . \frac{\partial }{\partial \rho }
E_{\rm RS2}(\rho,\lambda,Q,R) \right |_{(\rho,\lambda,R)=(1,1/2,R_b)} \cr
&&=\left . \frac{\partial}{\partial \rho}
\left \{
-\rho R_b \ln 2 -
\ln \left [ \sum_{y}
\left (\sum_{x}Q(x)P(y|x)^{\frac{1}{1+\rho}} \right)^{1+\rho}
\right ] \right \} \right |_{\rho=1}=0. 
\end{eqnarray}
This implies that for $R < R_b$,  
$\left . (\partial /\partial \rho )
E_{\rm RS2}(\rho,\lambda,Q,R) \right |_{\rho=1}
< \left . (\partial /\partial \rho )
E_{\rm RS1}(\rho,\lambda,Q,R) \right |_{\rho=1}$ holds
when the condition for a maximum is satisfied. 
Therefore, we should not select 
$E_{\rm RS1}(\rho,\lambda,Q,R)$, but rather $E_{\rm RS2}(\rho,\lambda,Q,R)$
for assessing the tightest bound $E_{\rm replica}(R)=
\mathop{\rm max}_{0 \le \rho,0 \le \lambda,Q}
\left \{E(\rho,\lambda,Q,R) \right \}$ for $R < R_b$, which yields
\begin{eqnarray}
E_{\rm replica}(R)&=&\mathop{\rm max}_{0 \le \rho,0 \le \lambda,Q}
\left \{E_{\rm RS2}(\rho,\lambda,Q,R) \right \} \cr
&=& \mathop{\rm max}_{1  \le \rho, 0 \le \lambda, Q}\left \{
-R \ln 2 -
\ln \left [ \sum_{y}
\left (\sum_{x}Q(x)P(y|x)^{1-\lambda \rho } \right)
\left (\sum_{x}Q(x)P(y|x)^{\lambda \rho } \right) \right ] \right \}\cr
&=& \mathop{\rm max}_{Q}\left \{
-R \ln 2 -
\ln \left [ \sum_{y}
\left (\sum_{x}Q(x)P(y|x)^{\frac{1}{2}} \right)^2
\right ] \right \}. 
\label{Ereplica2}
\end{eqnarray}
In the second line, any choice of $(\rho,\lambda)$ that satisfies 
$\lambda \rho =1/2$ and $\rho \ge 1$ optimizes the exponent. 

Although the style of the derivation seems somewhat different from that 
of the conventional approach, the exponents obtained by equations 
(\ref{Ereplica1}) and (\ref{Ereplica2}) are identical to 
those assessed using equation (\ref{randomcodingbound}). 
Therefore, $E_{\rm replica}(R)=E_r(R)$ holds, implying that 
no improvement is gained by the replica method
in the analysis of the ensemble of all codes.

Nevertheless, our approach is still useful for clarifying 
the origin of the seemingly artificial restriction $\rho \le 1$
in the conventional scheme. The above analysis indicates that 
there is no such restriction as long as 
the upper-bound of equation (\ref{second_bound}) is directly 
evaluated. Instead, what is the most relevant 
is the breaking of the analyticity with respect to $\rho$ of the upper-bound 
exponent $E(\rho,\lambda,Q,R)$, 
which can be interpreted as a phase transition between 
the two types of replica symmetric solutions
$E_{\rm RS1}(\rho,\lambda,Q,R)$ and
$E_{\rm RS2}(\rho,\lambda,Q,R)$ in the terminology of physics. 
As a consequence, we have to appropriately switch the functional forms of 
the objective function in order to 
correctly obtain the optimized exponent. 
However, this procedure, in practice, can be completely simulated by 
optimizing a single function in conjunction with introducing 
an additional restriction $\rho \le 1$, which can be summarized by 
a conventional formula of the random coding exponent,  
namely equation (\ref{randomcodingbound}). 

Of course, it must be kept in mind that the mathematical validity of 
our methodology is still open while the known correct 
results are reproduced. 
Although applying the saddle point assessment is 
a major reason for the weakening of mathematical rigor, 
the most significant issue in the current context 
is mathematical justification of the empirical 
criterion at $\rho=1$ to select the appropriate solution for $\rho \le 1$
when multiple saddle point solutions exist. 
Accumulated knowledge about error exponents of various 
codes in information theory 
\cite{Gallager1965,McElieceOmura1977,
Litsyn1999,Burnashev1984,Ashikhmin2000} may be of assistance 
for solving this issue. 

Although we have applied the replica method to 
an upper-bound following the conventional framework
in order to clarify the relation to an information theory method, 
it can be utilized to directly assess the minimum possible 
decoding error probability. 
For a region of lower $R$, there still exists a gap between 
the lower- and upper-bounds of the error exponents of 
the best possible code. 
An analysis based on the replica method indicates that 
the lower-bound of the exponent, which corresponds to 
the upper-bound of the decoding error probability, 
agrees with the correct solution \cite{Skantzos2003}.

\section{Analysis of low-density parity-check codes}
\subsection{Definition of an LDPC code ensemble}
Although a novel interpretation is obtained, 
our approach does not update known results in 
the analysis of the ensemble of all codes.	
However, this is not the case in general; 
the replica method usually offers 
a smaller upper-bound than conventional schemes for general code ensembles. 
We will show this for an ensemble of low-density 
parity-check (LDPC) codes. 

A $(k,j)$ LDPC code is defined by 
selecting $N-K$ parity checks composed of 
$k$ components, 
$x_{l_1}\oplus x_{l_2}\oplus \ldots \oplus x_{l_k}=0$, 
out of 
$
\left ( 
\begin{array}{c}
N \\
k 
\end{array} 
\right )
$
combinations of indices 
for characterizing a binary codeword of 
length $N$, $\bx=(x_l)\in \{0,1\}^N$, 
where $l_1,l_2,\ldots,l_k = 1,2,\ldots,N$ and 
$\oplus$ denotes addition over the binary field. 
There are several ways to define an LDPC code ensemble. 
For analytical convenience, we here focus on 
an ensemble constructed by 
uniformly selecting $N-K$ ordered combinations of 
$k$ different indices $l_1,l_2,\ldots,l_k$, $\paritycheck$, 
for parity checks, so that 
each component index of codewords $l(=1,2,\ldots,N)$ 
appears $j$ times in the total set of parity checks.
A code ${\cal C}$ constructed in this way is specified by a
set of binary variables $\bc=\{c_\paritycheck\}$, 
where $c_\paritycheck=1$ if the combination $\paritycheck$ is 
used for a parity check and $c_\paritycheck=0$ otherwise. 

For simplicity, we assume symmetric channels, where 
we can assume that the sent message $m$ is encoded into the null 
codeword $\bx=\bzero$. Under this assumption, the generalized
Chernoff's bound (\ref{orig_bound}) for an LDPC code is expressed as
\begin{eqnarray}
P_e({\cal C}) 
\le \sum_{\by}P(\by|\bzero)^{1-\lambda \rho}
\left (\sum_{\bx \ne \bzero }
{\cal I}(\bx|\bc)P(\by|\bx)^\lambda \right )^\rho, 
\label{LDPCchernoff}
\end{eqnarray}
where 
\begin{eqnarray}
{\cal I}(\bx|\bc)=\prod_{\paritycheck}
\left (1-c_\paritycheck+c_\paritycheck
\delta(x_{l_1}\oplus x_{l_2}\oplus \ldots \oplus x_{l_k},0)
\right ), 
\label{indicator}
\end{eqnarray}
returns unity if $\bx$ satisfies all the parity checks
and vanishes otherwise, screening only codewords in the summation over
$\bx$ in the right hand side of equation (\ref{LDPCchernoff}). 

\subsection{Performance assessment by the replica method}
Unlike the random code ensemble explored in the 
previous section, a statistical dependence arises
among codewords in an LDPC code. This yields atypically bad codes, 
the minimum distance of which is of the order of unity with 
a probability of algebraic dependence on $N$. 
The contribution of such atypical codes 
causes the average of the decoding error probability 
over a naive LDPC code ensemble to decay algebraically with respect to 
$N$, indicating that the error exponent vanishes
even for a sufficiently small rate $R$ \cite{MullerBurstein2001}. 
However, we can reduce the fraction of the bad codes to as small as
required by removing short cycles in the parity check dependence 
by utilizing certain feasible algorithms \cite{vanMourik2003}. 
This implies that, in practice, the performance of the LDPC code ensembles can be 
characterized by analysis with respect to the typical codes
utilizing the saddle point method as shown below \cite{KabashimaSaad2004}. 

In order to employ the replica method, we assess the average
of the right hand side of equation (\ref{LDPCchernoff}) with respect to 
the LDPC code ensemble 
\begin{eqnarray}
{\cal Q}(\bc)=\frac{1}{{\cal N}(k,j)}
\prod_{l=1}^N \delta \left (\sum_{\left \langle l_2 l_3 \ldots l_k \right 
\rangle}  c_{\left \langle l l_2 l_3 \ldots l_k \right 
\rangle},j \right ), 
\label{LDPCaverage}
\end{eqnarray}
where 
${\cal N}(k,j)=\sum_{\bc}
\prod_{l=1}^N \delta \left (\sum_{\left \langle l_2 l_3 \ldots l_k \right 
\rangle}  c_{\left \langle l l_2 l_3 \ldots l_k \right 
\rangle},j \right )$ stands for the number of $(k,j)$ LDPC codes. 
For $\rho=1,2,\ldots$ and sufficiently large $N$, 
evaluating this using the 
saddle method, substituting with $P(\by|\bx)=\prod_{l=1}^N
P(y_l|x_l)$, gives an upper-bound 
for the average decoding error probability 
over 
an ensemble of typical LDPC codes from which 
atypically bad codes are expurgated 
as $\overline{P_e({\cal C})} \le 
\exp \left [-N E_{\rm LDPC}(\rho,\lambda,R) \right ]$, 
where
\begin{eqnarray}
E_{\rm LDPC}(\rho,\lambda,R)
&=&-\mathop{\rm Extr}_{\chi, \widehat{\chi}}\left \{
\frac{N^{k-1}}{k!}
\sum_{\bb_1,\bb_2,\ldots,\bb_k} \prod_{t=1}^k \chi(\bb_t)
\prod_{a=1}^\rho \delta(b_{1}^a\oplus b_{2}^a\oplus \ldots \oplus b_{k}^a,0) 
\right . \cr
&+&\ln \left [\sum_{y}P(y|0)^{1-\lambda\rho}
\left (\sum_{\brx } 
\widehat{\chi}(\brx)^j \prod_{a=1}^\rho P(y|x^a)^\lambda \right )\right ] \cr
&-&\left . \sum_{\bb} \widehat{\chi}(\bb) \chi(\bb) 
-\left (\frac{j}{k}-j +j \ln \left [
\frac{(jN)^{1-1/k}}{\left( (k-1)!\right )^{1/k} } \right ] \right )
\right \}, 
\label{replica_exponent}
\end{eqnarray}
$\bb=(b^1,b^2,\ldots,b^\rho)\in \{0,1\}^\rho$ and 
$\brx=(x^1,x^2,\ldots,x^\rho)\in \{0,1\}^\rho$. 
$\mathop{\rm Extr}_{ X}$ denotes the operation of extremization 
with respect to $X$, which corresponds to the saddle point 
assessment of a certain complex integral and therefore 
does not necessarily mean maximization or minimization. 
An outline of the derivation is shown in Appendix A. 

An RS solution which is relevant for $0 \le \rho \le 1$
corresponding to RS1 in the previous section 
is obtained under the RS ansatz 
\begin{eqnarray}
\chi(\bb)&=& q \int_{-1}^{+1} d u \pi(u) \prod_{a=1}^\rho \left (
\frac{1+u (-1)^{b^a}}{2} \right ), \label{RSpi} \\
\widehat{\chi}(\bb)&=& \widehat{q} 
\int_{-1}^{+1} 
d \widehat{u} \widehat{\pi} (\widehat{u}) \prod_{a=1}^\rho \left (
\frac{1+\widehat{u} (-1)^{b^a}}{2} \right ), \label{RShatpi}
\end{eqnarray}
where $q$ and $\widehat{q}$ are normalization variables 
that constrain the respective variational functions 
$\pi(u)$ and $\widehat{\pi}(\widehat{u})$ to be distributions 
over $[-1,1]$, making it possible to analytically 
continue the expression (\ref{replica_exponent})
from $\rho=1,2,\ldots$ to $\rho \in \mR$. 
Carrying out partial extremization with respect to 
$q$ and $\widehat{q}$ yields an analytically continued
RS upper-bound exponent
\begin{eqnarray}
E_{\rm LDPC}^{\rm RS}
(\rho,\lambda,R)
\! &=& \!-\mathop{\rm Extr}_{\pi, \widehat{\pi}}\left \{
\frac{j}{k}
\ln \left [
\! \int_{-1}^{+1} \! \prod_{t=1}^k d u_t \pi(u_t) 
\left (\frac{1+\prod_{t=1}^k u_t}{2} \right )^\rho \right ]
\right . \cr
\! &+& \!\ln \left [\sum_{y}P(y|0)^{1-\lambda\rho}
\! \int_{-1}^{+1} \! \prod_{\mu=1}^j 
d\widehat{u}_\mu 
\widehat{\pi}(\widehat{u}_\mu )
\left (
\sum_{x=0,1} 
\prod_{\mu=1}^j
\left (\frac{1+\widehat{u}_\mu (-1)^x }{2}
\right )
P(y|x)^\lambda \right )^\rho \right ]\cr
\! &-& \! j \left . \ln \left [ \int_{-1}^{+1} 
du \widehat{\pi}(\widehat{u}) \pi(u) 
\left (\frac{1+\widehat{u}u }{2} \right )^\rho \right ] 
\right \}, 
\label{replica_exponent_continued}
\end{eqnarray}
where the functional extremization ${\rm Extr}_{\pi, \widehat{\pi}} \left \{
\cdots \right \}$ can 
be performed numerically in a feasible time by 
Monte Carlo methods in practice \cite{KabashimaSazuka2002}. 

\subsection{Comparison of lower-bound estimates of error threshold }
When the noise level is sufficiently small and 
the code length $N$ is sufficiently large, 
there exists at least one $(k,j)$ LDPC code
with a decoding error probability smaller than 
an arbitrary positive number. 
The maximum value of such noise levels is sometimes
termed the error threshold. 

Equation (\ref{replica_exponent_continued}) can be utilized to assess a 
lower-bound of the error threshold. 
Table \ref{table} shows the lower-bounds obtained by 
maximizing this equation with respect to $\rho\ge 0$ and $\lambda \ge 0$
for several sets of $(k,j)$. 
Estimates obtained by the conventional 
schemes utilizing Jensen's inequality, which in the current case 
are determined by an upper-bound exponent
\begin{eqnarray}
E_{\rm LDPC}^{\rm Jensen}
(\rho,\lambda,R)
\! &=& \!-\mathop{\rm Extr}_{u, \widehat{u}}\left \{
\rho \frac{j}{k}
\ln \left [
\left (\frac{1+u^k}{2} \right ) \right ]
\right . \cr
\! &+& \!\ln \left [\sum_{y}P(y|0)^{1-\lambda\rho}
\left (
\sum_{x=0,1} 
\left (\frac{1+\widehat{u}(-1)^x }{2}
\right )^j
P(y|x)^\lambda \right )^\rho \right ]\cr
\! &-& \! \rho j \left . \ln \left [ 
\frac{1+\widehat{u}u }{2} \right ] 
\right \}, 
\label{jensen_exponent}
\end{eqnarray}
are also provided for comparison. 

Table \ref{table} indicates that, in general, 
the lower-bounds estimated by the replica method are not smaller 
than those of the conventional schemes. 
This implies that unlike the case of the ensemble of all codes, 
employing Jensen's inequality can relax 
an upper-bound for general code ensembles and therefore 
there may be room for improvement in 
results obtained by conventional schemes based on this inequality.

\begin{table}[t]
\begin{center}
\begin{tabular}{|p{2cm}|p{2cm}|p{2cm}|p{2cm}|p{2cm}|p{2cm}|}
\hline
\centering $R$ &\centering $(j,k)$ & \centering 
Jensen 1& \centering Jensen 2& \centering replica & \parbox{2cm}{\centering Shannon}   \\
\hline 
\centering $1/2$& \centering $(3,6)$& \centering 0.0678&\centering 0.0915& \centering 0.0998 & 
\parbox{2cm}{\centering 0.109}  \\
\hline 
\centering $2/5$&\centering $(3,5)$ &\centering $0.115$ &\centering $0.129$ &	
\centering $0.136$ &\parbox{2cm}{\centering	$0.145$} \\
\hline 
\centering $1/3$ &\centering $(4,6)$ &\centering $0.1705$ & \centering
$0.1709$ & \centering $0.173$ &	\parbox{2cm}{\centering $0.174$} \\
\hline 
\centering $1/3$ &\centering $(2,3)$ &\centering $0$ &\centering $0.0670$ &
\centering $0.0670$ &\parbox{2cm}{\centering $0.174$}  \\
\hline
\centering $1/2$ &\centering $(2,4)$ &\centering $0$ & 
\centering $0.0286$ & \centering $0.0286$ & \parbox{2cm}{\centering $0.109$} \\
\hline 
\end{tabular}
\caption{Lower-bound estimates of the error threshold of BSC. 
In columns ``Jensen 1'', ``Jensen 2'' and ``replica'', 
the estimates represent the critical crossover rates $p_c$, below
which the maximized values of equation (\ref{replica_exponent_continued}) or 
(\ref{jensen_exponent}) are positive. 
In the evaluation, the exponents are maximized with respect to 
two parameters 
$\rho \ge 0$ and $\lambda \ge 0$ for ``Jensen 2'' and ``replica''
while a single parameter maximization with respect to $\rho \ge 0$,
keeping $\lambda=1/(1+\rho)$, is performed for ``Jensen 1''. 
``Shannon'' represents the channel capacity for a given code rate $R$. 
}
\label{table}
\end{center}
\end{table}

\section{Summary and discussion}
In summary, we have explored the relation between statistical mechanics and information theory methods
for assessing performance of channel coding, 
based on a framework developed 
by Gallager \cite{Gallager1965}. 
An average of a generalized Chernoff's bound for 
probability of decoding error over a given code ensemble 
can be directly evaluated by the replica method of statistical 
mechanics, while Jensen's inequality 
must be applied in a conventional information theory approach. 
The direct evaluation of the average 
associated a switch of two analytic functions 
in the random coding exponent known in information theory
with a phase transition between two replica symmetric solutions
obtained by the replica method. 
Better lower-bounds of the error threshold
were obtained for ensembles of LDPC codes under the assumption 
that the replica method produces the correct results. 
This may motivate an improvement in the accuracy of performance assessment,
refining the conventional methodologies. 

A characteristic feature of the methods developed in statistical 
mechanics is the employment of the saddle point assessment
utilizing a certain symmetry underlying the objective system, 
which, in some cases, makes it possible to accurately analyze 
macroscopic properties of large systems even when there are statistical 
correlations or constraints 
among system components. 
Such approaches may also be useful for analyzing codes of quantum information, 
for which, in many cases, there arise non-trivial correlations among codewords 
for the purpose of dealing with noncommutativity of operators 
\cite{Hayashi2006}. 

\ack
This work was partially supported by a Grant-in-Aid from MEXT, Japan, 
No. 1879006.  

\appendix
\section{Outline of derivation of equation (\ref{replica_exponent})}
Equation (\ref{replica_exponent}) is obtained by 
averaging the right hand side of equation (\ref{LDPCchernoff}) 
with respect to the $(k,j)$ LDPC code ensemble (\ref{LDPCaverage}). 
For this assessment, we first evaluate the normalization constant
${\cal N}(k,j)$ utilizing the identity
\begin{eqnarray}
\delta \left ( \sum_{\left \langle l_2l_3\ldots l_k \right \rangle}
c_{\left \langle l l_2l_3\ldots l_k \right \rangle}, j \right) 
=
\frac{1}{2\pi {\rm i}}\oint dZ_l Z_l^{-(j+1)} Z_l^{\sum_{\left \langle l_2l_3\ldots l_k \right \rangle}
c_{\left \langle l l_2l_3\ldots l_k \right \rangle}}, 
\label{contour}
\end{eqnarray}
where $\rm i=\sqrt{-1}$ and 
$\oint dZ$ denotes the contour integral along a closed curve 
surrounding the origin on the complex plane. 
Plugging this expression into ${\cal N}(k,j)=\sum_{\bc}
\prod_{l=1}^N \delta \left (\sum_{\left \langle l_2 l_3 \ldots l_k \right 
\rangle}  c_{\left \langle l l_2 l_3 \ldots l_k \right 
\rangle},j \right )$ yields
\begin{eqnarray}
{\cal N}(k,j)&=& \frac{1}{(2 \pi {\rm i})^N}
\oint \prod_{l=1}^N {d Z_l} Z_l^{-(j+1)}
\prod_{\left \langle l_1 l_2l_3\ldots l_k \right \rangle}
\left (1+Z_{l_1}Z_{l_2} \ldots Z_{l_k} \right ) \cr
&=&\frac{1}{(2 \pi {\rm i})^N}
\oint \prod_{l=1}^N {d Z_l} Z_l^{-(j+1)}
\exp \left [\sum_{\left \langle l_1 l_2l_3\ldots l_k \right \rangle}
\ln \left (1+Z_{l_1}Z_{l_2} \ldots Z_{l_k} \right ) \right ] \cr
&=&\frac{1}{(2 \pi {\rm i})^N}
\oint \prod_{l=1}^N {d Z_l} Z_l^{-(j+1)}
\exp \left [\sum_{\left \langle l_1 l_2l_3\ldots l_k \right \rangle}
Z_{l_1}Z_{l_2} \ldots Z_{l_k} + \mbox{higher order terms} \right ] \cr
&\simeq &\frac{1}{(2 \pi {\rm i})^N}
\oint \prod_{l=1}^N {d Z_l} Z_l^{-(j+1)}
\exp \left [\sum_{\left \langle l_1 l_2l_3\ldots l_k \right \rangle}
Z_{l_1}Z_{l_2} \ldots Z_{l_k} \right ] \cr
&\simeq &
\frac{1}{(2 \pi {\rm i})^N}
\oint \prod_{l=1}^N {d Z_l} Z_l^{-(j+1)}
\exp \left [\frac{N^k}{k!}
\left (\frac{1}{N}\sum_{l=1}^N Z_l \right )^k \right ]. 
\label{normalization}
\end{eqnarray}
Here, in the third to fifth lines we have omitted 
irrelevant higher order terms since they do not affect the following 
saddle point assessment. 
Inserting the identity 
$1=N^{-1}\int dq_0 \delta \left (\sum_{l=1}^NZ_l-N q_0 \right )
=(2 \pi N)^{-1}\int
dq_0 \int_{-{\rm i} \infty}^{+{\rm i}\infty}d \widehat{q}_0
\exp \left [ \widehat{q}_0 \left (\sum_{l=1}^NZ_l-N q_0 \right ) 
\right ]$ into this expression makes it possible 
to analytically integrate equation (\ref{normalization}) with respect
to $Z_l$ ($l=1,2,\ldots,N$). For large $N$, 
the most dominant contribution to the resulting integral 
with respect to $q_0$ and $\widehat{q_0}$ can be evaluated by the 
saddle point method as
\begin{eqnarray}
\frac{1}{N}\ln {\cal N}(k,j) &\simeq &
\mathop{\rm Extr}_{q_0, \widehat{q}_0}
\left \{
\frac{N^{k-1}}{k!} q_0^k -\widehat{q}_0q_0 
+\ln \left (\frac{\widehat{q}_0^j}{j!} \right )
\right \}=\frac{j}{k}-j+ \ln 
\left [
\frac{(jN)^{j-j/k}}{((k-1)!)^{j/k} j!} \right ], 
\end{eqnarray}
where the saddle point is given as 
$q_0=((k-1)!)^{1/k} j^{1/k}N^{-1+1/k}$ and 
$\widehat{q}_0=((k-1)!)^{-1/k}(jN)^{1-1/k}$. 

The average of the right hand side of equation (\ref{LDPCchernoff}) 
for $\rho=1,2,\ldots$ can be evaluated in a similar manner. 
For this, we expand $\left (\sum_{\bx \ne \bzero }
{\cal I}(\bx|\bc)P(\by|\bx)^\lambda \right )^\rho$ and 
take the average with respect to $\bc$, utilizing  
the LDPC code ensemble (\ref{LDPCaverage}). 
For each fixed set of $\bx^1,\bx^2,\ldots,\bx^\rho$, 
we obtain the expression
\begin{eqnarray}
&& 
\sum_{\bc}\delta \left ( \sum_{\left \langle l_2l_3\ldots l_k \right \rangle}
c_{\left \langle l l_2l_3\ldots l_k \right \rangle}, j \right) 
\prod_{a=1}^\rho {\cal I}(\bx^a|\bc) \cr
&&
=\frac{1}{(2 \pi {\rm i})^N} \oint \prod_{l=1}^N dZ_l 
Z_l^{-(j+1)}
\prod_{\left \langle l_1 l_2l_3\ldots l_k \right \rangle}
\left (1+Z_{l_1}Z_{l_2} \ldots Z_{l_k} 
\prod_{a=1}^\rho\delta(x_{l_1}^a\oplus x_{l_2}^a\oplus 
\ldots \oplus x_{l_k}^a,0)
\right ) \cr
&&\simeq \frac{1}{(2 \pi {\rm i})^N} \oint \prod_{l=1}^N dZ_l 
Z_l^{-(j+1)} 
\exp \left [
\sum_{\left \langle l_1 l_2l_3\ldots l_k \right \rangle}
Z_{l_1}Z_{l_2} \ldots Z_{l_k} 
\prod_{a=1}^\rho 
\delta(x_{l_1}^a\oplus x_{l_2}^a\oplus \ldots \oplus x_{l_k}^a,0 )\right ] \cr
&& \simeq \frac{1}{(2 \pi {\rm i})^N} \oint \prod_{l=1}^N dZ_l 
Z_l^{-(j+1)} \times \cr
&& \phantom{\exp }
\exp \left [
\sum_{\bb_1,\bb_2,\ldots,\bb_k}
\frac{N^k}{k!} \prod_{t=1}^k \left (\frac{1}{N} \sum_{l=1}^N Z_l 
\prod_{a=1}^\rho \delta(x_l^a,b_t^a) \right )
\prod_{a=1}^\rho \delta(b_1^a \oplus b_2^a \oplus \ldots \oplus b_k^a,0)
\right ], 
\label{decoupling}
\end{eqnarray}
where we have introduced the dummy variables 
$\bb_t=(b_t^1,b^2_t,\ldots,b^\rho_t)$ $(t=1,2,\ldots,k)$ as 
\begin{eqnarray}
\prod_{a=1}^\rho 
\delta(x_{l_1}^a\oplus x_{l_2}^a\oplus \ldots \oplus x_{l_k}^a,0)
=\sum_{\bb_1,\bb_2,\ldots,\bb_k}
\left (\prod_{a=1}^\rho \prod_{t=1}^k \delta(x_{l_t}^a,b_t^a)
\prod_{a=1}^\rho \right ) 
\delta(b_{1}^a\oplus b_{2}^a\oplus \ldots \oplus b_{k}^a,0), 
\end{eqnarray}
in order to decouple $x_{l_1}^a,x_{l_2}^a,\ldots, x_{l_k}^a$ 
of the left hand side. 
Inserting the identity 
\begin{eqnarray}
&&1=N^{-2^\rho} \int \prod_{\bb} d \chi(\bb) 
\delta \left (
\sum_{l=1}^N Z_l 
\prod_{a=1}^\rho \delta(x_l^a,b^a)-N \chi(\bx) \right ) \cr
&&=\frac{1}{(2 \pi N)^{2^\rho} }
\int \left (\prod_{\bb} d \chi(\bb) d \widehat{\chi}(\bb) \right )
\exp \left [  
\sum_{\bb} \widehat{\chi}(\bb)
\left (\sum_{l=1}^N Z_l 
\prod_{a=1}^\rho \delta(x_l^a,b^a)-N \chi(\bb) \right )
\right ] \cr
&&=\frac{1}{(2 \pi N)^{2^\rho} }
\int \left (\prod_{\bb} d \chi(\bb) d \widehat{\chi}(\bb) \right )
\exp \left [  
\sum_{l=1}^N Z_l 
\widehat{\chi}(\brx_l) 
-N\sum_{\bb} \widehat{\chi}(\bb)\chi(\bb) 
\right ], 
\end{eqnarray}
where $\brx_l=(x_l^1,x_l^2,\ldots,x_l^\rho)$ $(l=1,2,\ldots,N)$, 
into equation (\ref{decoupling}) allows 
integration with respect to $Z_l$ ($l=1,2,\ldots,N$) to be performed analytically. 
The resulting expression enables us to take summations with respect to 
$\brx_l$ ($l=1,2,\ldots,N$) independently in assessing the average, 
which yields identical contributions 
for $l=1,2,\ldots,N$ and leads to the saddle point 
evaluation of equation (\ref{replica_exponent}).

\end{document}